\documentclass[11pt]{article}

\usepackage{epsfig}

\newcommand{\be}{\begin{equation}}
\newcommand{\ee}{\end{equation}}
\newcommand{\bea}{\begin{array}}
\newcommand{\ea}{\end{array}}
\newcommand{\beqa}{\begin{eqnarray}}
\newcommand{\eeqa}{\end{eqnarray}}
\newcommand{\bean}{\begin{eqnarray*}}
\newcommand{\eean}{\end{eqnarray*}}

\def\up#1{\leavevmode \raise.16ex\hbox{#1}}

\def\sqr#1#2{{\vcenter{\vbox{\hrule height.#2pt
        \hbox{\vrule width.#2pt height#1pt \kern#1pt
          \vrule width.#2pt}
        \hrule height.#2pt}}}}

\newcommand{\gapproxeq}{\lower .7ex\hbox{$\;\stackrel{\textstyle >}{\sim}\;$}}
\newcommand{\lapproxeq}{\lower .7ex\hbox{$\;\stackrel{\textstyle <}{\sim}\;$}}

\def\thebibliography#1{{\bf REFERENCES\markboth
 {REFERENCES}{REFERENCES}}\list
 {[\arabic{enumi}]}{\settowidth\labelwidth{[#1]}\leftmargin\labelwidth
 \advance\leftmargin\labelsep
 \usecounter{enumi}}
 \def\newblock{\hskip .11em plus .33em minus -.07em}
 \sloppy
 \sfcode`\.=1000\relax}

\begin{document}

%\Large
\begin{center}
{\bf \large{ Classical  brackets for dissipative systems}}\\

\vspace{1cm}  Giuseppe Bimonte \footnote{Talk given by the
author.}, Giampiero Esposito, Giuseppe Marmo and Cosimo Stornaiolo
\end{center}
%\vspace{1cm}

\begin{center}
{\it   Dipartimento di Scienze Fisiche, Universit\`{a} di Napoli,
Federico II\\Complesso Universitario MSA, via Cintia, I-80126,
Napoli, Italy;
\\ INFN, Sezione di Napoli, Napoli, ITALY.\\
\small e-mail: \tt Bimonte@na.infn.it,
Giampiero.Esposito@na.infn.it,
Gimarmo@na.infn.it,  Cosmo@na.infn.it  } \\
\end{center}

\begin{abstract}
We show how to write a set of brackets for the Langevin equation,
describing the dissipative motion of a classical particle, subject
to external random forces. The method does not rely on an action
principle, and is based solely on the phenomenological description
of the dissipative dynamics as given by the Langevin equation. The
general expression for the brackets satisfied by the coordinates,
as well as by the external random forces, at different times, is
determined, and it turns out that they all satisfy the Jacobi
identity. Upon quantization, these classical brackets  are found
to coincide with the commutation rules for the quantum Langevin
equation, that have been obtained in the past, by appealing to
microscopic conservative quantum models for the friction
mechanism.

\end{abstract}

\section{Introduction}

The study of  dissipative systems is of fundamental interest in
many fields of physics, ranging from  statistical mechanics to
condensed matter, atomic physics etc.

It is well known that, classically, the action of a heath bath on
a particle can be described (say, in one space dimension) by the
"Langevin force": \be -\int_{-\infty}^t dt' \;\mu(t-t')\,\dot
x(t')\,+ F(t)\;, \ee where $\mu (t-t')$ is a friction coefficient,
called "memory function", and $F(t)$ is a random force. The memory
function is a phenomenological quantity, depending on the detailed
features of the coupling with the bath. Apart from the obvious
requirement that it should be non-negative at all times, it is
possible to show \cite{ford} that the second law of thermodynamics
requires its Fourier transform $\tilde \mu(z)$ to have a real
positive part on the real axis: \be {\rm Re} \;[\tilde \mu( \omega
+ i 0^+)] \ge 0 \;\;\;-\infty < \omega < \infty\;.\ee

The random force $F(t)$ is usually assumed to have zero mean:
$$ <F(t)>=0\;,$$ and
to satisfy the Gaussian property,  according to which all odd
correlators of $F(t)$ vanish, while the even ones can all be
written as sums of products of the two-point function $<F(t)
F(t')>$. The expression of the latter function, is determined by
the fluctuation-dissipation theorem, in terms of the temperature
$T$ of the bath, and of the memory function. For example, in the
case of a memory function of the form $\mu(t-t')=f \delta(t-t')$
(which corresponds to the original form of the Langevin equation),
one has: \be <F(t) F(t')>=2 k T f \,\delta(t-t')\;,\ee where $k$
is Boltzmann constant.

If the particle of mass $m$ is subject also to an external
conservative force, with potential $V(x)$, its motion is then
described by the (generalized) "Langevin equation": \be m \ddot x
+ \int_{-\infty}^t dt' \mu(t-t')\,\dot x(t')\,+\,V'(x)=F(t)\;. \ee

It is well known that this equation describes the approach to
equilibrium of the particle (if one assumes of course that the
heath bath is infinite in size and remains in equilibrium at all
times), and indeed one can prove that, for any choice of the
initial conditions, the probability density in the particle's
phase space approaches, for large times, the canonical
Maxwell-Boltzmann distribution.

It is natural to ask if an analogous  theory of dissipation
exists, for a quantum particle in interaction with a
quantum-mechanical bath. In particular, is there a quantum
Langevin equation, that one can use to describe the influence of
the bath on the quantum behavior of the particle? At a first
glance, it is not very clear what one should expect. Since the
Langevin force describes the influence of the bath on the
particle, and since the bath is itself a quantum system, it is
quite possible that both the random force $F(t)$ and the memory
function $\mu(t-t')$ become operators, in the quantum theory. The
simplest possibility, which we shall assume, is to keep
$\mu(t-t')$ a c-number. However, this is unlikely to be the case
for $F(t)$, since the l.h.s. of the Langevin equation, which
involves the particle's coordinate and momentum, is a-priori an
operator. In the Quantum theory, besides the randomness already
present in the classical theory (for $T>0$), there will exist a
further source of randomness due to intrinsic quantum
fluctuations. The latter should be encoded by a set of
commutators: \be [\hat F (t), \hat F(t')]\;\;\;,\;\;[\hat x(t),
\hat F(t')]\;\;\;,\;\;[\hat p(t), \hat F(t')].\ee Consideration of
unequal-time commutators appears to be necessary, because the
random force $\hat F (t)$ does not obey any equation of motion,
and so knowledge of equal-time commutators is not sufficient.

\section{A microscopic model for dissipation}

The standard approach to dissipation in Quantum Mechanics, is
based on the physical picture that dissipation arises from
coupling of the system of interest with a thermal bath. According
to this picture, one considers a $\it conservative$ microscopic
model for the bath, usually consisting of a large number of
oscillators,  and postulates a certain form for the system-bath
interaction. Elimination of the degrees of freedom of the bath
gives rise to an  effective equation of motion for the particle,
including both the damping force, and the fluctuating force. An
implementation of this philosophy, in the path-integral formalism,
was developed long ago by Feynman and Vernon \cite{feyn}.

In a canonical  framework, Ford et al. \cite{ford} proposed the
following simple independent oscillators model for the
particle-bath system: \be H=\frac{p^2}{2m}+V(x)+\sum_j
\left[\frac{p_j^2}{2 m_j}+ \frac{1}{2}m_j \omega_j^2
(q_j-x)^2\right] \;.\ee  One has the standard commutation rules:
\be [x,p]=i \hbar\;,\;\;\;[q_j,p_k]=i \hbar \delta _{jk}\;.\ee The
equations of motion implied by the above hamiltonian are: \be m
\ddot x+ V'(x)=\sum_j m_j \,\omega_j^2(q_j-x)\; \label{part}\ee
\be \ddot q_j +\omega_j^2 q_j=\omega_j^2 x \;.\label{oscil}\ee The
general solution of Eq.(\ref{oscil}) is: \be
q_j(t)=q_j^h(t)+x(t)-\int_{-\infty}^t
dt'\;\cos[\omega_j(t-t')]\dot x (t')\;, \label{sol}\ee where
$q_j^h(t)$ is the general solution of the homogeneous equation:
\be q_j^h(t)=q_j \cos(\omega_j t)+p_j \frac{\sin (\omega_j t)}{m_j
\omega_j}\;. \ee The choice of the retarded solution in
Eq.(\ref{sol}) explicitly breaks time-reversal. The picture one
has in mind  is that the particle is held fixed at $x=0$ at
$t=-\infty$. Upon plugging Eq.(\ref{sol}) into Eq.({\ref{part}),
we see that  the effective equation of motion for the particle of
interest becomes a Langevin Equation, with a memory function and a
fluctuating force, expressed in terms of the oscillator
parameters. In detail, one finds: \be \mu(t)=\sum_j m_j\,
\omega_j^2 \cos (\omega_j t)\Theta(t)\;, \ee where $\Theta(t)$ is
the Heaviside function. It is shown in \cite{ford} that, by a
suitable choice of parameters, one can reproduce the most general
form of the memory function, compatible with the principles of
thermodynamics. As for the random force, the model gives: \be
F(t)=\sum_j m_j \omega_j^2 q_j^h(t)\;, \ee and accordingly one
finds: \be \frac{1}{i \hbar}[F(t), F(t')]=- \sum_j m_j\,
\omega_j^3 \sin[\omega_j (t-t')]=\frac{d \mu}{dt}(t-t')+\frac{d
\mu}{dt}(t'-t)\;. \ee \noindent The question arises: how much of
these commutation relations depend on the detailed microscopic
model? A related question is whether it is possible to obtain the
commutators for $x(t)$, $p(t)$ and $F(t)$ directly from the
Langevin Equation, without making recourse to microscopic models?

\section{PB for dissipative systems}

Our approach to the quantization of the Langevin's equation
proceeds in a way similar to that  followed for conservative
systems:

\begin{itemize}
    \item Step 1: define first a set of Poisson
Brackets (PB) $\{\;,\;\}$ on the classical "phase space" ${\cal
P}$;
    \item  Step 2: replace PB's by commutators:
$$\{A\,,B \} \rightarrow \frac{1}{i \hbar}\,[\hat A\,, \hat
B\,]\;\;.
$$
\end{itemize}

\noindent The problems  that arise in step 1, are twofold:
\begin{itemize}
    \item Problem 1: what is the particle's phase space $\cal P$? Since the particle
is acted on by a random force, initial data do not determine its
future evolution. In fact, since $F(t)$ can a priori be anything,
any path $x(t)$ whatsoever is a possible trajectory for the
particle. Thus, if we think of the phase space as the set of
actual motions of the particle, we need to take for ${\cal P}$ the
{\it infinite dimensional} set of all paths $x(t)$:
$$
{\cal P}=\{x: \bf R \rightarrow \bf R\}
$$
    \item Problem 2: dissipative equations cannot be
derived, in general, from an action principle. How does one get
then a PB on ${\cal P}$?
\end{itemize}
In order to address both questions, we have elaborated
\cite{bimonte} a definition of PB's, that relies {\it directly and
exclusively on Langevin's equation}. To see how this can be done,
let us look at the conservative limit, when $\mu(t)=F(t)=0$: \be m
\ddot x + V'(x)=0\;.\ee Then, a canonical Poisson bracket exists,
and we can use it to define a retarded two-point function:
$$ G^-(t,t'):=\{x(t), x(t')\}\;,{\rm for}\;\; t \ge t'\;,$$ \be
G^-(t,t')\equiv 0\;,\;{\rm for}\;\; t < t'\;,\ee We can easily
derive the equation satisfied by $G^-(t,t')$, for $t>t'$. Indeed,
linearity of the PB, and the Leibnitz rule imply: $$ m
\frac{d^2}{d t^2}G^-(t,t')=m \frac{d^2}{dt^2}\{x(t), x(t')\}=m
\{\ddot x(t), x(t')\}=$$ \be=-\{V'(x(t)),
x(t')\}=-V''(x(t))\{x(t),x(t')\}=-V''(x(t))G^-(t,t')\;.\ee   \ In
view of the boundary condition: \be \lim_{t\rightarrow
t'}m\frac{d}{dt}G^-(t,t')=\{m\dot x(t),x(t)\}=-1\;,\ee we can
write an equation valid for all times:\be \left(m \frac{d^2}{d
t^2}+V''(x(t))\right)G^-(t,t')=-\delta(t,t')\;,\ee which shows
that $G^-(t,t')$ is the {\it retarded} Green's function for small
perturbations of the classical motion $x(t)$. One can similarly
define an advanced two-point function $G^+(t,t')$: $$
G^+(t,t'):=-\{x(t), x(t')\}\;,{\rm for}\;\; t \le t'\;,$$ \be
G^+(t,t')\equiv 0\;,\;{\rm for}\;\; t > t'\;.\ee It is easy to
verify that $G^+(t,t')$ is the advanced Green's functions for
small perturbations of classical motions. If we define the
"commutator function" $\tilde G (t,t')$: \be \tilde G (t,t')
\equiv G^-(t,t')-G^+(t,t')\;,\ee we obviously have: \be
\{x(t),x(t')\}=\tilde G (t,t')\;\;\;\;\;\;\;\forall \;\;t,
t'.\label{peier}\ee We see that antisymmetry of the PB follows
from the {\it reciprocity relation}: \be
G^+(t,t')=G^-(t',t)\;.\label{rec}\ee

\noindent Our proposal \cite{bimonte} to define the PB for the
Langevin equation, is to take Eq.(\ref{peier}) as the {\it
definition} of the PB. This way of defining PB, in terms of the
Green's functions associated with the operator that describes
small perturbations of classical motions, was introduced, for
systems admitting an action principle, by Peierls \cite{peierls}
(For a review of Peierls brackets, see Refs. \cite{dewitt,
bimo2}.) . Here, we are trying to extend this method to
non-conservative systems, for which an action principle cannot be
found in general.
 So, we consider the retarded Green's function for the Langevin
equation: \be \left(m \frac{d^2}{d t^2}+V''(x(t))\right)G^-(t,t')+
\int_{-t'}^t
d\tau\,\mu(t-\tau)\frac{d}{d\tau}G^-(\tau,t')=-\delta(t-t')\;.\label{pert}\ee
Notice that the lower extremum of integration in the driving term
is $t'$, because of the retarded boundary condition. Notice also
that the "background" curve $x(t)$ is an {\it arbitrary} path,
since any path is a possible motion for the particle, the random
force being arbitrary.

\noindent We take $G^+(t,t')$ as the Green's function of the {\it
adjoint} of Eq.(\ref{pert}): \be \left(m \frac{d^2}{d
t^2}+V''(x(t))\right)G^+(t,t')- \int_{t}^{t'}
d\tau\,\mu(\tau-t)\frac{d}{d\tau}G^+(\tau,t')=-\delta(t-t')\;.\label{adj}\ee

\noindent With this choice, the reciprocity relation
Eq.(\ref{rec}) still holds. This is easily seen if we write
Eqs.(\ref{pert}) and (\ref{adj}) in the form of integral equations
(with singular kernels): \be \int _{-\infty}^{\infty} d \tau
\,L(t,\tau)G^-(\tau,t')=-\delta(t-t') \ee \be \int
_{-\infty}^{\infty} d \tau
\,L^T(t,\tau)G^+(\tau,t')=-\delta(t-t')\;, \ee where the
superscript $T$ denotes transpose (transpose is the same as
adjoint, because we are in the real field). We switch to DeWitt
\cite{dewitt} condensed notation, in which the continuous time
variable is treated as a discrete index and so linear
integro-differential operators are written as matrices. In this
notation: \be
L_{ij}G^{-jk}=-\delta_i^k\;\;\;,\;\;\;(L^T)_{ij}G^{+jk}=-\delta_i^k\;.\label{green}\ee
Multiplication of the second equation by $G^{-il}$ gives: \be
G^{-il}(L^T)_{ij}G^{+jk}=G^{-il} \delta_i^k=G^{-kl}\;.\ee However,
using the first of Eq.(\ref{green}), the l.h.s. is also equal to:
\be G^{-il}(L^T)_{ij}G^{+jk}=G^{-il}L_{ji}G^{+jk} =\delta_j^l
G^{+jk}=G^{+lk}\;.\ee Comparison of the r.h.s. of the above two
Eqs. proves the reciprocity relation.

\noindent The important issue is to check the Jacobi identity. A
direct computation gives: \be \{\{x^i,x^j\},x^k\}+{\rm cycl.
perm}=\tilde G ^{il} \tilde G ^{jk}_{\;\;,\,l}+ \tilde G ^{jl}
\tilde G ^{ki}_{\;\;,\,l}+ \tilde G ^{kl} \tilde G
^{ij}_{\;\;,\,l}\equiv {\cal T}^{ijk}\;,\ee and so the Jacobi
identity is fulfilled is the quantity ${\cal T}^{ijk}$ vanishes.
Use of reciprocity relations allows to write ${\cal T}^{ijk}$ only
in terms of $G^{-ij}$, and its functional derivatives. The latter
are easily obtained from Eq.(\ref{green}): \be
L_{ij,l}G^{-jk}+L_{ij}G^{-jk}_{\;\;\;\;\;\;,\;l}=0\;.\ee
Multiplication by $G^{+mi}$, and use of reciprocity relation,
gives: \be
G^{-mk}_{\;\;\;\;\;\;\;,\,l}=-G^{-im}L_{ij,l}G^{-jk}\;.\ee A
simple computation then gives: \be {\cal
T}^{ijk}=(G^{-li}G^{-mj}G^{-nk}+ c.p.)(L_{mn}-L_{nm})_{,\;l}\ee
This shows that the Jacobi identity holds if and only if the
antisymmetric part of the operator for the perturbations, $L$, has
vanishing functional derivatives (namely is  independent on the
unperturbed path $x(t)$). This is surely the case if the equations
of motion derive from an action, because then $L_{nm}$ is the
second functional derivative of the action $S$, and since
derivatives commute, it is symmetric: \be
L_{mn}=S_{,mn}=S_{,nm}=L_{mn}\;.\ee Now, the operator $L$ for the
Langevin equation is not symmetric, however its antisymmetric part
has vanishing functional derivatives, because of the linear
character of the friction term. Indeed, if we restore the plain
notation for integral operators, we can write the antisymmetric
part of $L$ (acting on a function $\psi(t)$) as: \be
(L-L^T)\psi(t)=\int_{-\infty}^{\infty}
dt'\;[\mu(t-t')+\mu(t'-t)]\dot \psi(t')\ee The kernel of $L-L^T$
is independent on the unperturbed path $x(t)$ and so it has null
functional derivatives. Then, the friction term does not spoil the
Jacobi identity.

\noindent By taking the Langevin Equation as a definition of
$F(t)$, we can evaluate the PB satisfied by the external force:
\be \{F(t), F(t')\}=\frac{d \mu}{dt}(t-t')+\frac{d
\mu}{dt}(t'-t)\;. \ee This is the same as ($1/(i \hbar)$) times
the commutator that was obtained  in \cite{ford}, by using an
explicit microscopic model. \noindent It is also possible to
verify that the equal-time PB of the particle's coordinate and
velocity remain canonical: $$ \{x(t),x(t)\}=\{\dot x(t), \dot
x(t)\}=0\;,$$
$$ m \{x(t), \dot x(t)\}=1\;.$$

\noindent  It is important to observe that, when friction is
present, $F(t)$ has non trivial brackets, and so it is {\it
inconsistent} to set $F(t)=0$. This means that a consistent PB can
be written for the Langevin equation, only if the particle is
acted on by an external force. However, if the friction term is
zero, $F(t)$ has vanishing brackets with everything (including the
particle's coordinate); then, it is possible to take $F(t)=0$ and
to restrict the bracket onto the set of solutions of the equations
of motion. In this way we recover back the (finite dimensional)
phase space of the conservative system, and its canonical PB.

\section{Conclusions}

We have defined a set of PB for the generalized Langevin equation.
Distinctive features of our approach are:
\begin{itemize}
    \item the method does not require an action principle, and is
    based directly on the equations of motion;
    \item it can be applied to dissipative equations, for which an action principle does
    not exist, like the Langevin equation. In this way, we obtained the PB directly from
    the macroscopic description of dissipation,
    as provided by the Langevin equation, without making recourse
    to microscopic models;
    \item when dissipation is present, the relevant phase space
    is, a priori, the space of all paths, which is an
    infinite-dimensional space;
    \item in the absence of dissipation, when the system is
    conservative, it is possible to set the random force to zero
    and restrict the bracket onto the standard phase space, spanned by
    classical solutions of the e.o.m.. One recovers then the usual
    canonical PB.
\end{itemize}

G.B., G.E. and G.M. acknowledge partial financial support by PRIN
2002 {\it SINTESI}.


\begin{thebibliography}{200}

\bibitem{ford} G.W. Ford, J.T. Lewis and R.F. O'Connell, Phys.
Rev. {\bf A 37}, 4419 (1988).

\bibitem{feyn} R.P. Feynmnan and F.L. Vernon Jr., Ann. Phys. {\bf 24}, 118 (1963); ibid. {\bf
281}, 547 (2000).


\bibitem{bimonte} G. Bimonte, G. Esposito, G. Marmo and C.
Stornaiolo, New classical  brackets for dissipative systems,
hep-th/0303206.

\bibitem{peierls} R. E. Peierls, Proc. Roy. Soc. (London), {\bf A 214},
143 (1952).

\bibitem{dewitt} B.S. DeWitt, {\it Dynamical Theory of Groups} {\it and
Fields}, (Gordon \& Breach, New York, 1965).

\bibitem{bimo2} G. Bimonte, G. Esposito, G. Marmo and C.
Stornaiolo, Int. J. Mod. Phys. {\bf A 18}, 2033 (2003).


\end{thebibliography}
\end{document}